\documentclass[floats,twocolumn,prl,shortbibliography,10pt]{revtex4-2}
\usepackage{amsfonts,amsmath,graphicx,epsfig,amssymb}
\usepackage{epsfig}
\usepackage{mathrsfs}
\usepackage{latexsym}
\usepackage{amsxtra}
\usepackage{amsbsy}
\usepackage{amscd}
\usepackage{color}
\usepackage{bbm}
\usepackage{hyperref}
\usepackage{multirow}
\usepackage{graphics}
\usepackage{amsfonts}
\usepackage{subfigure}
\usepackage{color}
\allowdisplaybreaks

\newcommand{\sectitle}[1]{\vspace{.5cm}{\em #1.--}}

\begin{document}

\title[]{New insights into the oscillations of the nucleon electromagnetic form factors}
\author{Qin-He Yang$^{1,2}$}
\email{yqh@hnu.edu.cn}
\author{Di Guo$^{1,2}$}
\email{diguo@hnu.edu.cn}
\author{Ling-Yun Dai$^{1,2}$}
\email{Corresponding author: dailingyun@hnu.edu.cn}
\author{Johann Haidenbauer$^{3}$}
\email{j.haidenbauer@fz-juelich.de}
\author{Xian-Wei Kang$^{4,5}$}
\email{xwkang@bnu.edu.cn}
\author{Ulf-G. Mei{\ss}ner$^{6,3,7}$}
\email{meissner@hiskp.uni-bonn.de}
\affiliation{$^{1}$ School of Physics and Electronics, Hunan University, Changsha 410082, China}
\affiliation{$^{2}$ Hunan Provincial Key Laboratory of High-Energy Scale Physics and Applications, Hunan University, Changsha 410082, China}
\affiliation{$^{3}$ Institute for Advanced Simulation, Institut f\"ur Kernphysik and J\"ulich Center for Hadron Physics, Forschungszentrum J\"ulich, D-52425 J\"ulich, Germany}
\affiliation{$^{4}$ Key Laboratory of Beam Technology of the Ministry of Education, College of Nuclear Science and Technology, Beijing Normal University, Beijing 100875, China}
\affiliation{$^{5}$ Institute of Radiation Technology, Beijing Academy of Science and Technology, Beijing 100875, China}
\affiliation{$^{6}$ Helmholtz Institut f\"ur Strahlen- und Kernphysik and Bethe Center for Theoretical Physics, Universit\"at Bonn, D-53115 Bonn, Germany}
\affiliation{$^{7}$ Tbilisi State University,Tbilisi 0186, Georgia}

\begin{abstract}
The electromagnetic form factors of the proton and the neutron in the timelike region are
investigated. Electron-positron annihilation into antinucleon-nucleon ($\bar NN$)
pairs is treated in distorted wave Born approximation, 
including the final-state interaction in the $\bar NN$ system. The latter is obtained by a
Lippmann-Schwinger equation 
for $\bar{N}N$  potentials derived within SU(3) chiral effective field theory. 
By fitting to the phase shifts and (differential) cross section data, a high quality description is achieved. With these amplitudes, the oscillations of the electromagnetic form factors of 
the proton and the neutron are studied. It is found that each of them can be described by two fractional oscillators. One is characterized as \lq overdamped' and dominates near the threshold, while the other is \lq underdamped' and plays an important role in the high-energy region. These two oscillators are essential to understand the distributions of polarized electric charges induced by hard photons for the nucleons.


\end{abstract}

\maketitle

\sectitle{Introduction}\label{Sec:I}
The electromagnetic form factors (EMFFs) of the nucleons are an important topic in nuclear and particle physics. 
They parameterize the nucleon's response to a virtual photon and play a crucial role in exploring the
nucleon structure. The EMFFs in the timelike region are accessible in
the process of electron-positron annihilation into antinucleon-nucleon pairs. For  reviews, see e.g. Refs.~\cite{Denig:2012by,Pacetti:2014jai}. 
Over the last decade, quite a few measurements have been
reported~\cite{BaBar:2013ves,CMD-3:2015fvi,CMD-3:2018kql,Achasov:2014ncd,Druzhinin:2019gpo,BES:2005lpy,BESIII:2015axk,BESIII:2019tgo,BESIII:2019hdp,BESIII:2021rqk,BESIII:2021tbq,SND:2022wdb}
and the uncertainties have been greatly reduced as compared to
older data~\cite{Bardin:1994am,Castellano:1973wh,Antonelli:1993vz,Antonelli:1994kq,Antonelli:1998fv,Delcourt:1979ed,Bisello:1983at,DM2:1990tut}. 
One interesting aspect of these improved measurements was that they revealed an apparent oscillatory behavior of the proton EMFFs 
\cite{BaBar:2013ves,BESIII:2019hdp}. This behavior, first strongly emphasized in ~\cite{Bianconi:2015owa}, 
has been the subject of many other theoretical studies since then 
\cite{Lorenz:2014yda,Bianconi:2015vva,Bianconi:2016bss,Tomasi-Gustafsson:2020vae,Lin:2021umz,Lin:2021xrc,Cao:2021asd,Dai:2021yqr,Qian:2022whn,Tomasi-Gustafsson:2022tpu}. 
However, unlike the clear meaning of the EMFFs in the spacelike region \cite{Punjabi:2015bba,Perdrisat:2006hj,Cloet:2013jya},  a plausible and generally accepted explanation for this remarkable and unexpected feature is still missing. 
Very recently, the BESIII Collaboration~\cite{BESIII:2021tbq} provided the first high-statistics measurement of the neutron EMFFs.
Interestingly, also an oscillatory behavior was observed over the 
measured energy range of $2.0-3.2$~GeV. Analyzing the data, the BESIII Collaboration 
found that there should be a phase difference in oscillations of the EMFFs between the proton and the neutron. 

In the present work, we want to shed new light on these oscillations by considering both the proton and neutron EMFFs. 
In particular, we want to extend the study of the oscillations to the low-energy region. 
Measurements around the thresholds are difficult, 
and the analysis of the EMFFs in Ref.~\cite{BESIII:2021tbq} has been restricted to the energy region 
above $2$~GeV. 
Meanwhile, there has been a strong interest in the near-threshold region, both by theorists and experimentalists. 
There are clear enhancements around the thresholds, not only for 
$e^+e^-\to \bar NN$ (antinucleon-nucleon pair), but also in electron-positron annihilation into other antibaryon-baryon pairs, 
see e.g. Refs.~\cite{Dai:2017fwx,BESIII:2017kqg,BESIII:2017hyw}, which can be attributed to the effects of the final-state interaction (FSI) as in \cite{Yao:2020bxx,Lin:2021umz}. 
Accordingly, we take into account the $\bar NN$ FSI to obtain reliable predictions for the 
EMFFs around the $\bar pp$ and $\bar nn$ thresholds. The implemented FSI effects are based on an $\bar NN$ interaction derived within chiral effective field theory (ChEFT).
For a detailed analysis of the behavior of the EMFFs, we use fractional oscillation functions to fit the data and our predictions from ChEFT, from the $\bar NN$ thresholds 
to $3.2$~GeV. 









\sectitle{Formalism}\label{Sec:II}
ChEFT provides a systematic way to deal with the dynamics of the nucleon-nucleon interaction 
in the low-energy region \cite{Epelbaum:2008ga,Machleidt:2011zz}. This approach has also
been successfully extended to studies of the
$\bar NN$ interaction \cite{Kang:2013uia,Dai:2017ont}. To cover the energy region up to $2.2$~GeV, 
we consider $\bar{N}N$  potentials up to next-to-leading order (NLO) within SU(3) ChEFT. 
The $\bar{N} N$ scattering amplitudes are obtained by solving the Lippmann-Schwinger equation \cite{Kang:2013uia,Dai:2017ont}.  
Based on these amplitudes, we construct the $e^+e^-\to \bar NN$ amplitudes in the framework of the distorted wave  Born approximation (DWBA) \cite{Haidenbauer:2014kja,Dai:2017fwx,Dai:2018tlc,Haidenbauer:2020wyp}, 
where the FSI in the $\bar NN$ system is taken into account\footnote{ 
	Note that this framework has also been applied to study EMFFs of other baryons, specifically to  
	$e^+e^-\to \bar{\Lambda}\Lambda$ \cite{Haidenbauer:2016won},
	$e^+e^-\to\bar\Sigma\Sigma, \, \bar \Xi \Xi$ \cite{Haidenbauer:2020wyp}, 
	$e^+e^-\to \Lambda_c^+ \bar{\Lambda}_c^-$ \cite{Dai:2017fwx}. }. 
Then we evaluate the EMFFs of nucleons from threshold up to $2.2$~GeV. 

The differential cross section of $e^+ e^-\to \bar{N}N$ can be written in terms of the 
EMFFs as \cite{Haidenbauer:2014kja}
\begin{align}\label{eq:dsigma2}
	\frac{\mathrm{d}\sigma}{\mathrm{d}\Omega}&=\;\frac{\alpha^2\beta}{4s}C(s)\left[|G^N_{\mathrm{M}}(s)|^2(1+\cos^2\theta) \right.\nonumber\\
	&\;\;\;\;\;\;\left.+\frac{4M_N^2}{s}|G^N_{\mathrm{E}}(s)|^2\sin^2\theta\right],
\end{align}
where $\beta$ is a phase space factor, $\beta=k_N/k_e$ with $k_N$, $k_e$ the three-momenta of nucleon, electron in the center-of-mass frame, $s$ is the Mandelstam variable, 
$s=4\, (M^2_N + k^2_N)$, and $C(s)$ is the Sommerfeld-Gamow factor \cite{Haidenbauer:2014kja}.  
The cross section is obtained from Eq.(\ref{eq:dsigma2}) by integration. 

For a reliable description of the energy dependence of the reaction amplitude, the 
FSI in the $\bar{N}N$ system should be included. 
This is done within the DWBA. Accordingly, one has the formula for the $\gamma\bar{N}N$
vertices \cite{Haidenbauer:2014kja,Dai:2017fwx}
\begin{align}
	f^{\bar{N}N}_{L}(k;E_k)&=f^{\bar{N}N,0}_{L}(k)+\sum_{L'}\int_0^{\infty}\frac{dpp^2}{(2\pi)^3}f^{\bar{N}N,0}_{L'}(p)\nonumber\\
	&\times\frac{1}{2E_k-2E_p+i0^+}T_{L'L}(p,k;E_k), \label{eq:DWBA}
\end{align}
where $E_k=\sqrt{s}/2=\sqrt{k^2+M^2_N}$. $f^{\bar{N}N,0}_{L}(k)$ 
is the Born term (or bare $\gamma\bar{N}N$ vertex) and contains two constants, 
$G^{p,0}_{\mathrm{E}}$ and $G^{n,0}_{\mathrm{E}}$.
They are complex due to the contributions of inelastic channels, such as 
$\gamma\to \pi \pi\to \bar{N}N$.  
The $\bar{N}N$ scattering amplitude $T_{LL'}(p,p';E_k)$ for the $^3S_1$-$^3D_1$ coupled partial
waves is solved from the Lippmann-Schwinger equation (LSE)
\begin{align}
	&T_{L''L'}(p'',p';E_k)=V_{L''L'}(p'',p')+\nonumber\\
	&\sum_L\int\frac{dpp^2}{(2\pi)^3}V_{L''L}(p'',p)\frac{1}{2E_k-2E_p+i0^+}\nonumber\\
	&\hspace{3.5cm} \times T_{LL'}(p,p';E_k),  \label{eq:LS}
\end{align}
where $p',\, p,\, p''$ is the three-momentum of the initial, intermediate and final $\bar{N}N$ states,
respectively. $V_{L''L}(p'',p')$ is the $\bar{N}N$ interaction potential calculated within SU(3) ChEFT up to NLO, including one/two pseudoscalar boson exchanges (OBE/TBE) and 
contact terms. The OBE/TBE potentials can be obtained via the $G$-parity 
transformation from the relevant $NN$ potentials \cite{Haidenbauer:2013oca,Dai:2017ont},   
the contact terms are given by~\cite{Dai:2017ont,Epelbaum:2014efa,Kang:2013uia}, 
and the annihilation part is parameterized by an unitarity approach \cite{Kang:2013uia,Dai:2017ont}.
As in Refs.~\cite{Epelbaum:2004fk,Haidenbauer:2013oca,Dai:2017fwx}, the LSE is regularized by an exponential regulator, $f_R(\Lambda)=\exp[-(p^6+p'^6)/\Lambda^6]$.
To explore the sensitivity to the choice of cut-offs, we employed a range of values, 
i.e., $\Lambda=[750-950]$~MeV, with steps of $50$~MeV,  which all led to similar results.
The quantitatively best description was achieved with $\Lambda=850$~MeV. 
For simplicity, we will only show results for that cutoff here.  

\begin{figure}[t!]
	\centering
	\includegraphics[width=0.99\linewidth,height=0.16\textheight]{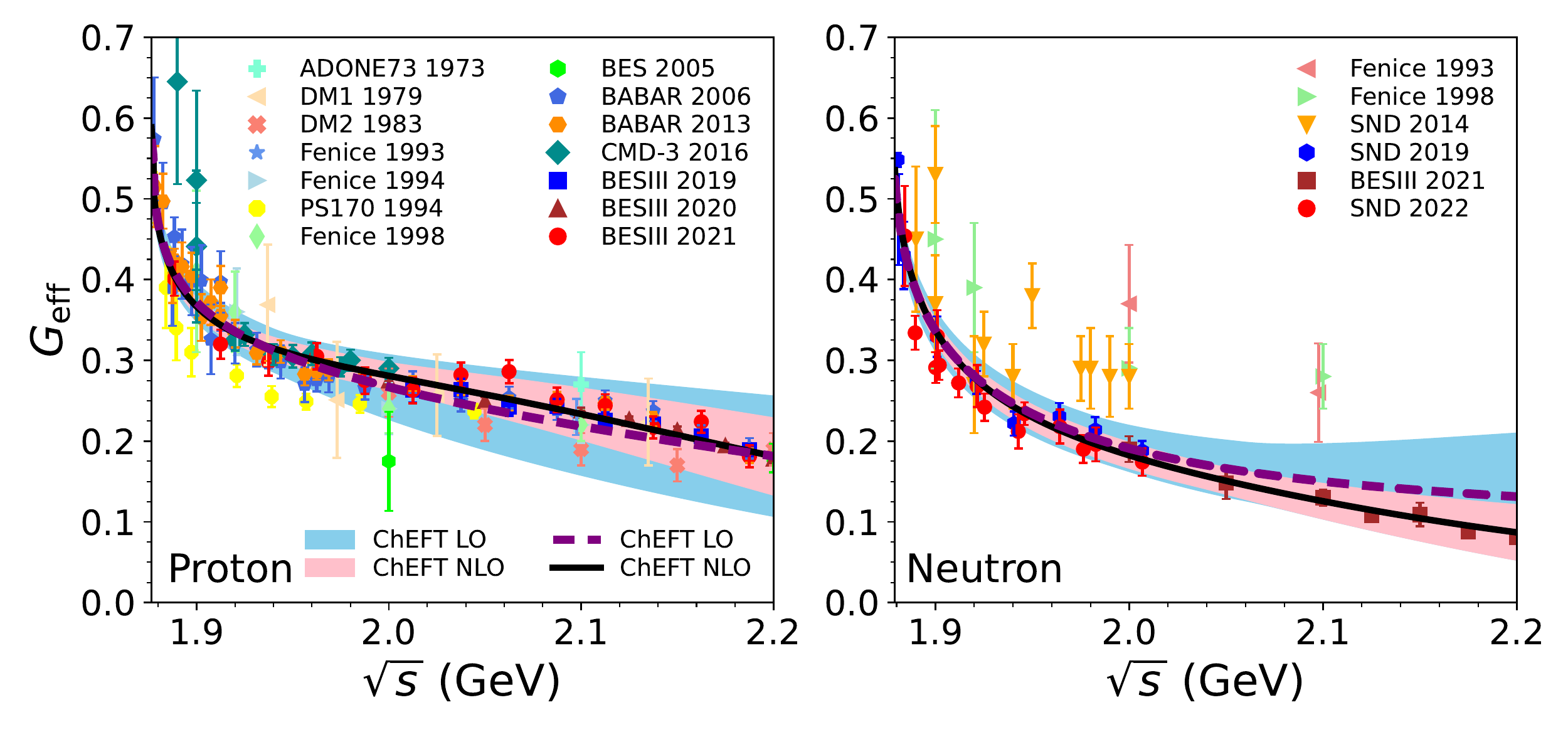}
	\caption{(Color online) Results for the effective EMFFs. The data points are 
		from ADONE73~\cite{Castellano:1973wh}, 
		Fenice~\cite{Antonelli:1993vz,Antonelli:1994kq,Antonelli:1998fv}, DM1~\cite{Delcourt:1979ed}, 
		DM2~\cite{Bisello:1983at}, BABAR~\cite{BaBar:2005pon,BaBar:2013ves}, 
		CMD-3~\cite{CMD-3:2015fvi,CMD-3:2018kql}, 
		BESIII~\cite{BES:2005lpy,BESIII:2019hdp,BESIII:2019tgo,BESIII:2021rqk,BESIII:2021tbq}, 
		SND~\cite{Achasov:2014ncd,Druzhinin:2019gpo,SND:2022wdb}, and PS170~\cite{Bardin:1994am}. 
		\label{Fig:effective} }
\end{figure}

\sectitle{Results and discussion}\label{Sec:III}
The low-energy constants (LECs) and values for $G^{p,0}_{\mathrm{E}}$ and $G^{n,0}_{\mathrm{E}}$
are fixed by a combined fit to the $\bar NN$ phase shifts 
and scattering lengths, as well as to $e^+e^-\leftrightarrow \bar{N}N$ (differential) 
cross sections and EMFFs.  
Regarding the $\bar{N}N$ data, only the phase shifts given by the PWA~\cite{Zhou:2012ui} in the very low-energy region are considered. 
The parameters $G^{N,0}_{E}$ are mainly determined by fitting to 
the $e^+e^- \to \bar NN$ data.   
With the $e^+e^-\to \bar NN$ amplitudes fixed, we can predict the EMFFs. 
Here we consider the effective EMFFs, which are defined by 
\begin{equation}
	|G^N_{\mathrm{eff}}(s)|=\sqrt{\frac{\sigma_{e^+e^-\to \bar{N}N}(s)}{\frac{4\pi\alpha^2\beta}{3s}C(s)
			[1+\frac{2M_N^2}{s}]}}\, , ~ (N=n,p)~,\label{eq:Geff} 
\end{equation}
as there are few differential cross section data that would allow for a separation of the two complex EMFFs.
The results for $|G^p_{\mathrm{eff}}|$ and $|G^n_{\mathrm{eff}}|$ are shown in 
Fig.~\ref{Fig:effective}.
The uncertainty is estimated following Refs.~\cite{Epelbaum:2014efa,Dai:2017ont}, generated by one class of Bayesian naturalness priors \cite{Furnstahl:2015rha}.
As can be seen, the effective EMFFs drop off rapidly right from the $\bar NN$ threshold and 
then decrease more slowly with increasing $\sqrt{s}$. 
The effective EMFF of the neutron is a bit smaller than that of the proton. 
This may be caused by the fact that the net charge of valence quarks of the neutron is zero.

To study the oscillatory behavior of the EMFFs suggested by the data in
Refs.~\cite{BaBar:2013ves,BESIII:2021tbq},  
we introduce subtracted form factors (SFFs) by subtracting the dipole 
contribution \cite{Bianconi:2015owa,BESIII:2021rqk,BESIII:2021tbq} 
\begin{equation}
	G^N_{\mathrm{osc}}(s)=|G^N_{\mathrm{eff}}(s)|-G^N_D(s)\,, \label{eq:G;osc}
\end{equation}
where $G_D$ is the dipole expression for the nucleon~\cite{Bianconi:2015owa,BESIII:2021rqk}.
Combining our ChEFT amplitudes up to $2.2$~GeV and the data sets at higher energies, a complete description of the SFFs from the threshold 
up to $3.2$~GeV is possible.

Then what is the underlying physics behind these SFFs?
It is found that the fractional oscillation functions \cite{2001Dynamics} can
fit the SFFs rather well~\footnote{Indeed, our model even describes the data well up to 5~GeV. However, the data there has significant uncertainties, so that we do not consider them  here.}. 
Notice that with ordinary oscillators of exponential and trigonometric functions (see, e.g., Ref.~\cite{Bianconi:2015owa}), one can not describe the oscillation well from threshold up to 2~GeV, as there is a substantial enhancement near the threshold. 
We suggest
\begin{align}
	G^{N}_{\mathrm{osc}}(\tilde{p})&=G_{\mathrm{osc},1}^{0,N}(0)\tilde{E}_{\alpha_1^{N},1}(-\omega_1^2 \tilde{p}^{\alpha_1^{N}}) \nonumber\\
	&+G_{\mathrm{osc},2}^{0,N}(0)\tilde{E}_{\alpha_2^N,1}(-\omega_2^2(\tilde{p}+p_0^N)^{\alpha_2^N}) \,, \label{eq:fraction}
\end{align}
with the Mittag-Leffler function $\tilde{E}_{\alpha,\beta}(z)$ given by~\cite{1956Higher}.
The subscripts $1,2$ are for two kinds of oscillators. Here, $\tilde{p}$ is the momentum of
the antinucleon in the rest-frame of the nucleon \cite{Bianconi:2015owa},
$\tilde{p}\equiv\sqrt{E_N^2-M_{N}^2}$,  $\quad E\equiv s/(2M_{N})-M_{N}$. 
The momentum shift ($p_0^N$) describes the \lq phase delay' of the 
oscillation between the proton and neutron for the second oscillator. 
Consequently, one can set $p_0^p=0$ and  $p_0^n\neq 0$, where the latter will be fixed by a fit. 
$\omega_{1,2}$ are the oscillation frequencies 
for the two oscillators, and we set each of them to be the same for proton and neutron, 
as inspired by Ref.~\cite{BESIII:2021tbq}. This is reasonable since the proton and 
neutron are isospin doublets.   
$\alpha_{1,2}^{N}$ are the damping factors. When $\alpha=1$, one has  
$\tilde{E}_{1,1}(z)=e^{z}$, and the fractional oscillation will restore to a 
normal \lq overdamped' oscillation. It is also required that $\alpha_2^p=\alpha_2^n$ 
to ensure that the oscillations of proton and neutron are the same, but only the 
phase and modulus are different. 
$G_{\mathrm{osc},1,2}^{0,N}(0)$ are initial values for the two different oscillators. 
As shown in Eq.~(\ref{eq:G;osc}), they are combined together and we have $G_{\mathrm{osc}}^{0,p}(0)=0.1918$ and 
$G_{\mathrm{osc}}^{0,n}(0)=0.3138$ given by ChEFT. Hence, we only need two independent initial values. 
Specifically, the equations of motions of the fractional oscillators are given by 
\begin{align}
	G_{\mathrm{osc}}^{N}(\tilde{p})&=G_{\mathrm{osc},1}^{N}(\tilde{p})+G_{\mathrm{osc},2}^{N}(\tilde{p}) ,\nonumber\\	
	G_{\mathrm{osc},j}^{N}(\tilde{p})&=G_{\mathrm{osc},j}^{0,N}-\frac{\omega_j^2}{\Gamma(\alpha_{j}^{N})}\int_0^{\tilde{p}+p_0^N}(\tilde{p}+p_0^N-t)^{\alpha_{j}^{N}-1} \nonumber\\
	& \hspace{3.5cm} \times G_{\mathrm{osc},j}^{N}(t)dt  \,.     \label{eq:G;motion} 
\end{align}
where the subscripts \lq $j$=1,2' are the two oscillators. The fit parameters are shown in Fig.\ref{tab:fraction}.
\begin{table}[ht]
	\centering
	\begin{tabular}{c|cc}
		\hline
		Parameters         &    proton          & neutron         \\
		\hline
		$\beta(10^{-2})$   & 6.020$\pm$0.034 &17.453$\pm$0.023 \\
		$\alpha_1$         &1.263$\pm$0.002  &1.060$\pm$0.001 \\
		$\alpha_2$         &1.880$\pm$0.001  &1.880$\pm$0.001 \\
		$\omega_1(10^{-2})$&5.371$\pm$0.015  &5.371$\pm$0.015 \\
		$\omega_2(10^{-3})$&7.472$\pm$0.022  &7.472$\pm$0.022 \\
		$p_0$\,(MeV)       &   0             &1035.93$\pm$2.44\\
		\hline
	\end{tabular}
	\caption{The fit parameters of the fractional oscillators. Notice that $\alpha_2$ and $\omega_2$ of underdamped oscillators are the same for the proton and neutron. The errors of the parameters are taken from MINUIT.}
	\label{tab:fraction}
\end{table}

\begin{figure}[t!]
	\centering
	\includegraphics[width=0.99\linewidth,height=0.25\textheight]{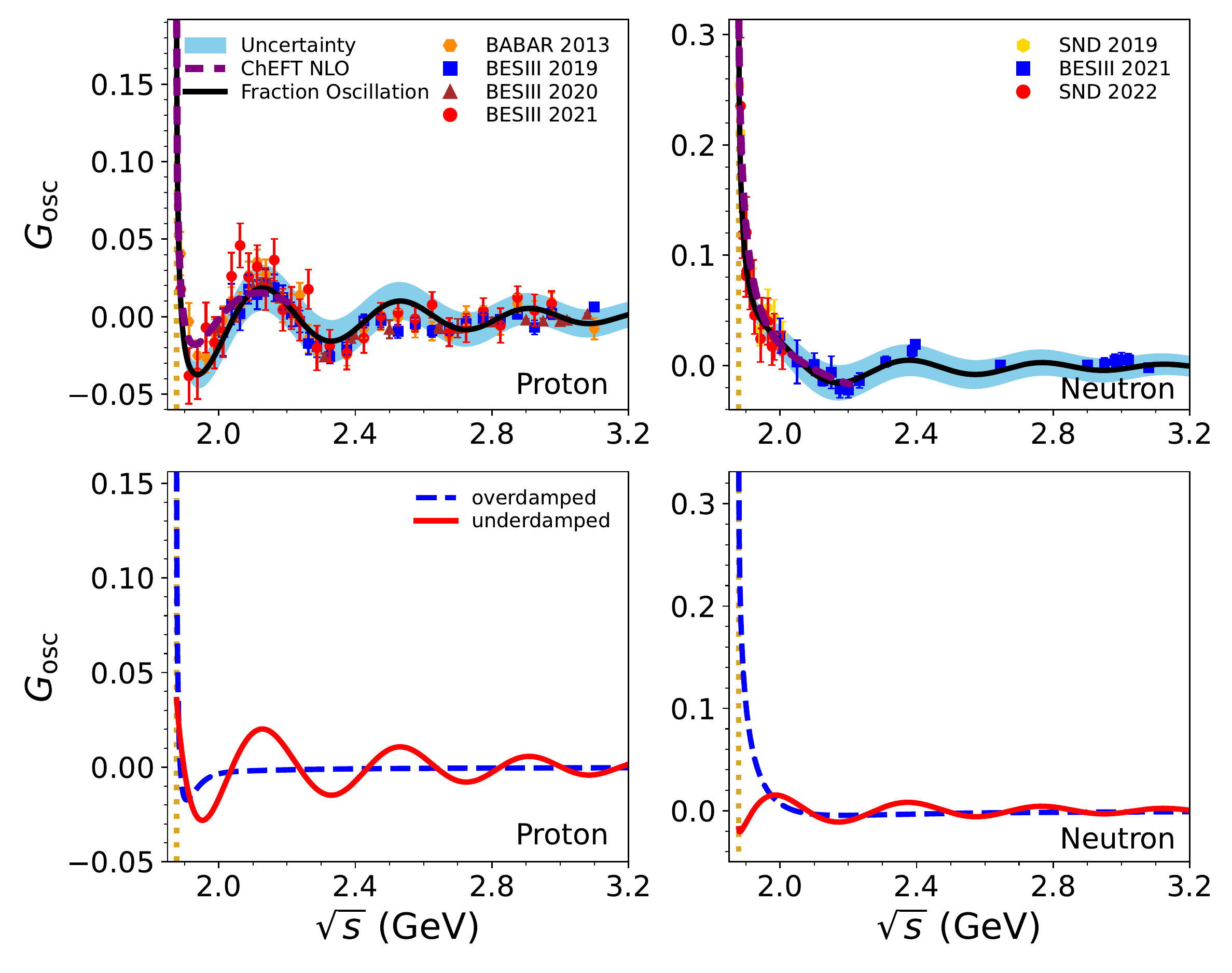}
	\caption{(Color online) Results for the SFFs $G_{\mathrm{osc}}^{N}(\tilde{p})$ with fractional oscillation functions. See Eqs.~(\ref{eq:G;osc},\ref{eq:fraction}). 
		The yellow dotted lines are the thresholds. The uncertainty bands are estimated from bootstrap \cite{Efron:1979bxm}, within 1~$\sigma$. 
	}
	\label{fig:GoscFO1}
\end{figure}

The SFFs based on 
the very recent data sets \cite{BaBar:2013ves,Druzhinin:2019gpo,BESIII:2021rqk,BESIII:2021tbq,SND:2022wdb} 
and our predictions from ChEFT   
are described rather well with two fractional oscillators, see the graphs in Fig.~\ref{fig:GoscFO1}. 
The data points of Refs.~\cite{BESIII:2019tgo,BESIII:2019hdp} are also superimposed 
for the reader's convenience.
The first term in Eq.~\eqref{eq:fraction} dominates the oscillation behavior around the
threshold, and then it decreases rapidly with increasing energy. Its oscillation is
similar to an \lq overdamped vibration', where the magnitude decreases quickly without obvious fluctuations. Indeed we have $\alpha_1^p=1.26$ and $\alpha_1^n=1.06$, close to one. 
Therefore, we call this an \lq overdamped' oscillator.   
The second term ($\alpha_2^p=\alpha_2^n=1.88$) describes a much slower decreasing oscillatory behavior and dominates in the high-energy region, named as \lq underdamped' oscillator.  

To see clearly the contributions of each oscillator, we draw the individual
contributions at the bottom of Fig.~\ref{fig:GoscFO1}. 
As can be seen, the \lq underdamped' oscillators start from different positions for the proton and the neutron, while they have the same period function, corresponding to a translation/delay on the momentum/energy. 
This confirms the \lq phase delay' proposed in Ref.~\cite{BESIII:2021tbq}. 
For the \lq overdamped' oscillators, that of the proton still shows \lq oscillation' around the threshold, whereas that of the neutron only decreases. 
This is consistent with the fixed damping factors, $\alpha^p_1>\alpha^n_1$. Further, they dominate in the low-energy region.

Compared with the regular oscillator proposed in Refs.~\cite{Bianconi:2015owa,BESIII:2021tbq},
our fractional oscillators can not only describe well the oscillation behavior above 2~GeV,
but also in the energy region close to the threshold.  
Interestingly, the fractional oscillators
describe the data in the energy region of [1.9, 2.0]~GeV even better than  ChEFT,
as shown in Fig.~\ref{fig:GoscFO1}. 
This confirms the reliability of our model. 
Further, the EMFFs also have an \lq overdamped' behavior around the
threshold, which has not been realized before. It shows  that the FSI of the antinucleon-nucleon
pair very near the threshold should be strong.

The fractional oscillation equation (with $1<\alpha_j^N <2$)  is in the middle of two limits: diffusion ($\alpha_j^N=1$) and wave 
equations ($\alpha_j^N=2$). The diffusion solution of a multi-particle system is caused by nonuniform distributions of, for example, density, 
while a wave usually moves with a constant period in a uniform medium. 
Our fractional oscillators reveal the distributions of the higher order polarized electric charges for the nucleons \footnote{For the relation between vacuum polarization and the timelike EMFFs, see Supplementary material. Also, the LO polarization effect is of course due the dipole.}, which can be 
divided into two parts:  The \lq quadrupole' contribution from the underdamped oscillation,  also called  uniform distributions for simplicity; And the \lq octupole' contribution from the overdamped oscillation, called nonuniform distributions.

We perform the Fourier transformation on the EMFFs to study the distributions ($\mathcal{D}_{\rm eff}^{N}(r)$) of the 
polarized electric charges for the nucleons, including individual contributions of the dipoles and the oscillators. 
One has \cite{Bianconi:2015owa}
\begin{align}
	\mathcal{D}_{\rm eff}^{N}(r)=\frac{1}{(2\pi)^3}\int d^3\vec{r}~ G_{\rm eff}^{N}(\tilde{p})~\exp(-i\vec{\tilde{p}}\cdot\vec{r}) .    \label{eq:D;fourier} 
\end{align}
Here, $r$ is the distance between the polarized electric charges and the nucleons. 
\begin{figure}[t!]
	\centering
	\includegraphics[width=0.99\linewidth,height=0.25\textheight]{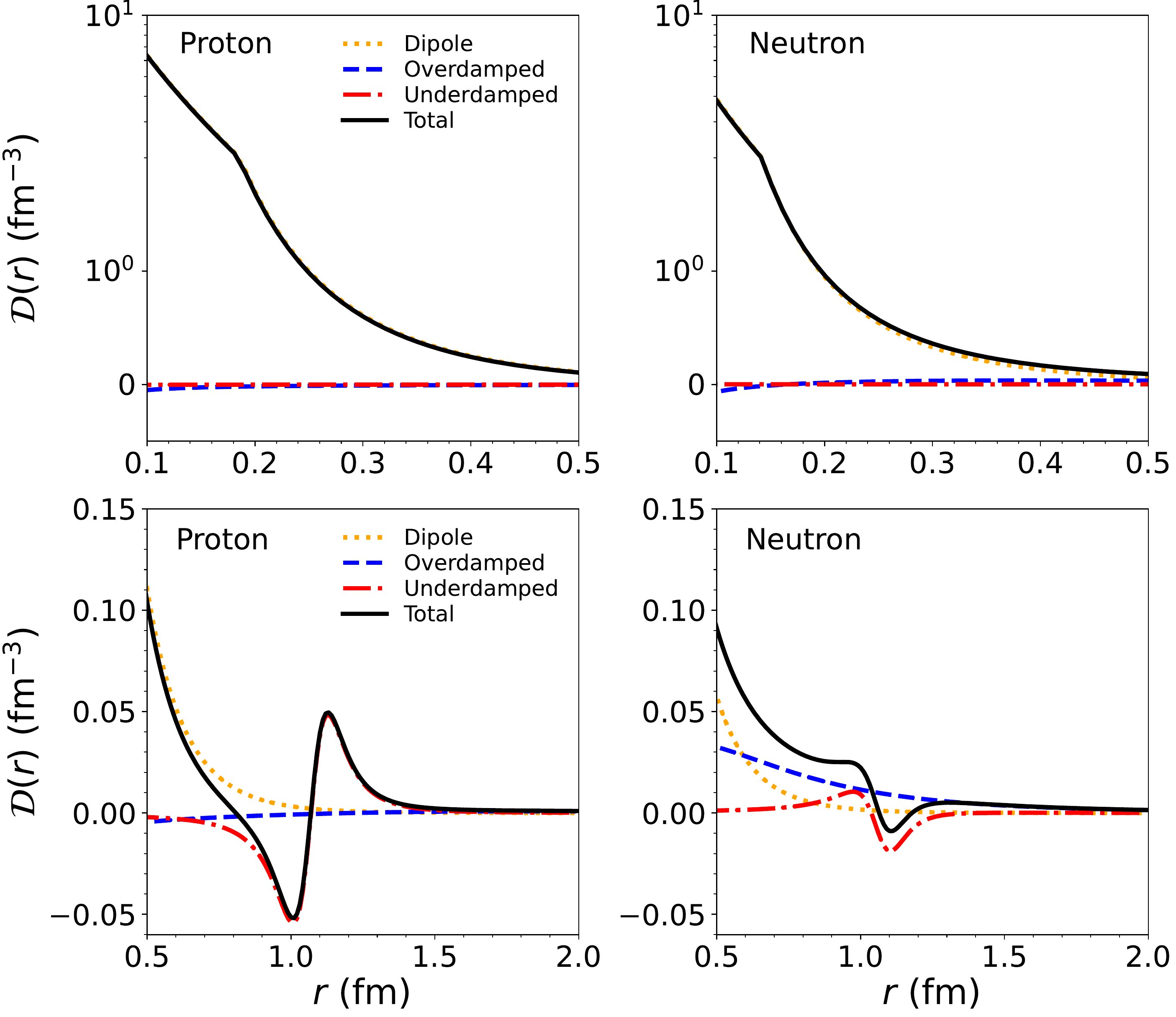}
	\caption{(Color online) The Fourier transformation of the EMFFs in the range of $r\in [0.1, 2]$~fm. 
		\label{fig:Fourier} }
\end{figure}
As shown in Fig.~\ref{fig:Fourier}, the dipole contribution is almost the same as that of the total one at short distances: 
$r<0.5$~fm for the proton and $r<0.2$~fm for the neutron. The difference between them in the range 
$0.2<r<0.5$~fm for the neutron is noticeable, as we lack enough precise data to fix the solutions (and the dipole formula).  
The overdamped and the underdamped contributions are much smaller. 
In the long-distance region, all of them  are rather small.  

In the middle range of $0.5-1.5$~fm the total $D(r)$'s have a wave shape around $1$~fm, the typical length
scale of the strong interactions. This is due to the underdamped oscillators, as shown in Fig.~\ref{fig:Fourier}.
More specifically, when $r$ decreases, the distributions of polarized electric charges of the proton would climb 
to a positive peak and then fall to a negative trough (positive out side and negative inside, the same as charge screening), while the
opposite situation 
occurs for the neutron. This is because the proton has $uud$ valence quarks, while the neutron contains $udd$. 
The vacuum polarization would be enhanced as $r$ decreases. Hence, for the proton the polarized charges would be increased in the positive direction. 
However, for the neutron, there are more $d$ quarks, and their polarization is stronger than the $u$ quarks. Thus, the net polarized charges are not 
zero but enhanced in the negative direction. Also, the polarization of $u$ quarks will cancel parts of the $d$ quarks, so it has smaller magnitudes 
for the \lq peak' and \lq trough', compared with that of the proton.  
When $r$ continues to decrease, the anti-screening of gluons would drive the polarized quarks to be free, and the uniform distributions 
will be restored gradually.  The discussions above also explain the phase difference in the 
oscillations between the proton and neutron, where the peaks are shifted by roughly $0.1$~fm. 
As $r$ decreases from $1$ to $0.5$~fm, the distributions of the overdamped oscillators will decrease to negative values or rise in the 
positive direction for the proton and the neutron, respectively. This is consistent with the fact that the nonuniform distributions are from higher order polarization effects and are opposite to the uniform one.  

To test the stability of our fractional model, we also use other functions for the background to obtain the SFFs, and then fit our fractional oscillators to them. 
It is found that the oscillation is still apparent, and our fractional model describes the new SFFs well, see the discussion in the Supplementary material. 
The dynamics of the fractional oscillation model should be further studied in the future. 

To study the origin of the oscillation, we use the potential of ChEFT described above
in the low-energy region ($2m_p-2.2$~GeV), 
and the relativistic potential of one-gluon exchanges (OGEs) between quark and antiquark 
pairs \cite{Godfrey:1985xj} in the high energy region ($2.2-3.2$~GeV) 
with only constituent quarks, $uud$ for the proton and 
$\bar{u}\bar{u}\bar{d}$ for the antiproton considered. This is compatible with the fact that there are not many sea quarks, and 
leads to the perturbative QCD behavior of the
EMFFs. Also, we do not consider OGEs between internal quarks of a nucleon/antinucleon. 
Taking the potentials into Eqs.(\ref{eq:DWBA},\ref{eq:LS}), we obtain the amplitudes and extract the EMFFs. Finally, we get the SFFs. 
They have similar behavior to the fractional oscillations. Especially, the overdamped oscillation is dominated by the ChEFT potential, 
and the underdamped one is mainly caused by the $s-$channel OGEs. This gives clear clues about how the oscillation is generated. 
See Supplementary material for details.   

\sectitle{Summary}\label{Sec:IV}
In this Letter, we investigate the EMFFs of the nucleons in the timelike region. 
The FSI in the antinucleon-nucleon system has been taken into account 
in distorted wave Born approximation, 
based on an $\bar NN$ interaction derived within SU(3) ChEFT up to NLO.
The experimental data of (differential) cross
sections of $e^+e^-\to \bar{N}N$ as well as the phase shifts of $\bar{N}N$ scattering 
are fitted to fix the free parameters. 
A high-quality description of the $e^+e^-\to \bar NN$ data has been obtained.

A more detailed analysis of the EMFFs of the nucleon 
suggested that the oscillations seen for the proton and the neutron
should be considered as a combined effect of two fractional oscillators.  
One is \lq overdamped',  dominating near the threshold, implying 
strong FSI effect close to the threshold. The other one is \lq underdamped',
dominating in the high-energy region, which also confirms the \lq phase delay' of oscillations
between the proton and the neutron.   
Our model of two fractional oscillators can describe well the EMFFs from threshold up to
3.2~GeV, which eventually sheds new light on the inner structure and dynamics of the nucleon: The \lq underdamped' oscillation is caused by uniform distributions of higher order polarized electric charges induced by hard photons for the nucleons, which also generates the phase difference in the oscillations between the proton and the neutron. The \lq overdamped' oscillation is caused by the nonuniform distributions. With a combined potential of ChEFT and the relativistic quark model, the fractional oscillation can be reproduced.

\sectitle{Acknowledgments}\label{Sec:V}
We thank the helpful discussions with Professors Z. W. Liu and Q. F. L\"u about the quark model. 
This work is supported by Joint Large Scale Scientific Facility Funds of the National
Natural Science Foundation of China (NSFC) and Chinese Academy of Sciences (CAS) under Contract No.U1932110, NSFC Grants No.11805059, 11805012, 11675051, 12322502, and 12335002, and Fundamental Research Funds for the central Universities.
It was further supported by Deutsche Forschungsgemeinschaft (DFG) and NSFC through funds provided to the Sino-German CRC 110 ``Symmetries and the Emergence of Structure in QCD" (NSFC
Grant No.~11621131001, DFG Grant No.~TRR110).
The work of UGM was supported in part by VolkswagenStiftung (Grant no. 93562)
and by the CAS President's International Fellowship Initiative (PIFI) (Grant No.~2018DM0034).

\bibliography{ref}

\appendix
\setcounter{equation}{0}
\setcounter{table}{0}
\setcounter{figure}{0}
\renewcommand{\theequation}{\Alph{equation}}
\renewcommand{\thetable}{\Alph{table}}
\renewcommand{\thefigure}{\Alph{figure}}

\onecolumngrid
\section{Supplemental material}
\sectitle{Vacuum polarization for the nucleon}
The timelike EMFFs of nucleons are obtained through the processes of the $e^+e^-\to\gamma^* \to \bar{N}N$ or $\bar{N}N \to\gamma^* \to e^+e^-$. The momenta in the center-of-mass frame (CMF) of nucleon and the anti-nucleon are given as   
\begin{eqnarray}
p_{e^+}=(\sqrt{s}/2,-\vec{p}_e)\,,\, p_{e^-}=(\sqrt{s}/2,\vec{p}_e)\,,
p_{\gamma^*}=(\sqrt{s},0) \,, p_{\bar{N}}=(\sqrt{s}/2,-\vec{p}_N)\,,\,p_{N}=(\sqrt{s}/2,\vec{p}_N) \,. \nonumber
\end{eqnarray}
Notice that by crossing symmetry, the out-going/in-going antinucleon is equal to an in-going (out-going) nucleon with minus momentum $-p_{\bar{N}}$.
To see the underlying physics of the timelike EMFFs, we change from the CMF to the rest-frame of the nucleon. One has 
\begin{eqnarray}
   p_{N, in}=(m_N,0)\,,\,p_{N, out}=(-E_N,-\vec{\tilde{p}}~) \,, \,  p_{\gamma^*,out}=(\frac{s}{2m_N},\vec{\tilde{p}}~) \,,  \nonumber
\end{eqnarray}
with $\vec{\tilde{p}}$ and $E_N$ as defined in the main text.  
In the timelike region, the energy range of $E_N$ is $[m_N,\infty)$ and that of the momentum $\tilde{p}$ is $[0,\infty)$. The latter implies that the 
Fourier transformation is performed in the whole momentum space and it is complete.  
It indicates that 
a nucleon at rest emits a hard virtual photon (with energy above $2m_N$ and momentum $\vec{\tilde{p}}$) and finally its energy becomes negative. Of course, by crossing symmetry, it also indicates a nucleon (with negative energy) absorbs a hard virtual photon,  
\begin{eqnarray}
p_{N, in}=(-m_N,0)\,,\, p_{\gamma^*,in}=(\frac{s}{2m_N},\vec{\tilde{p}}~)
\,,\,p_{N, out}=(E_N,\vec{\tilde{p}}~) \,. \nonumber
\end{eqnarray}
Inserting the $\gamma^* N \bar{N}$ vertex into the left graph of Fig.~A, one finds that  the vacuum polarization around the nucleon is recovered, 
see the discussion below. 
\begin{figure}[h!]
    \centering
    \includegraphics[width=0.8\linewidth,height=0.16\textheight]{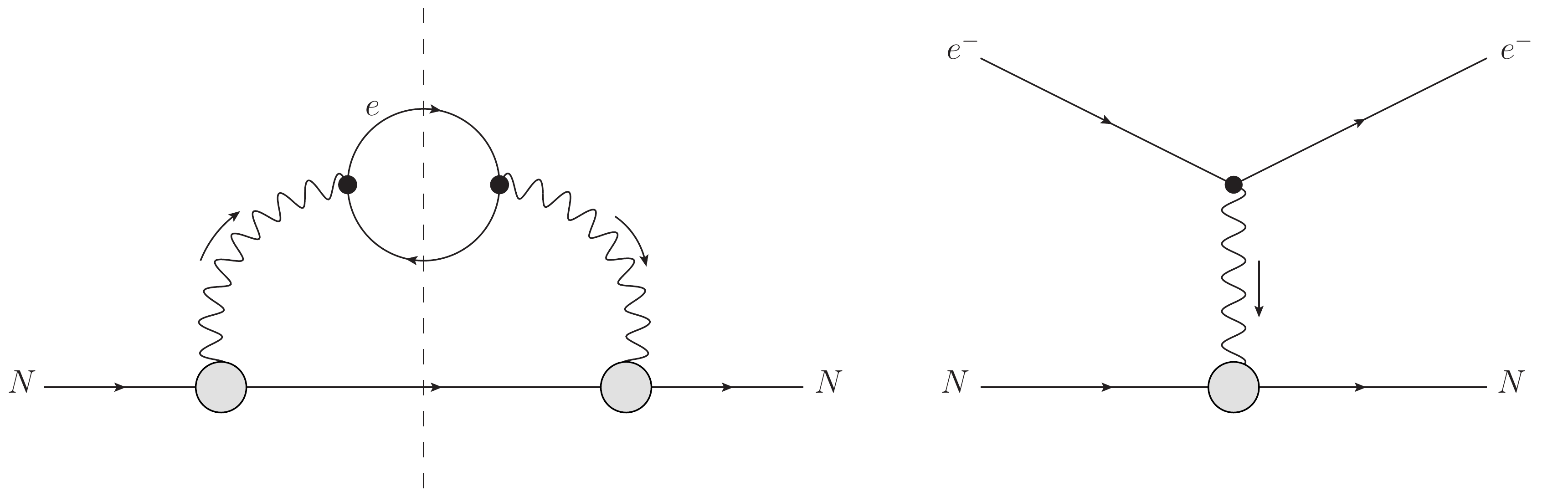}
    \caption{Feynman diagrams to illustrate the charge screening (on the left side) and the elastic $e-p$ scattering in the spacelike region.  
    \label{fig:TL_SL} }
\end{figure}
The radius $r$ introduced by the Fourier transformation of Eq.~(10) indicates the distance between the virtual photon and 
the nucleon at rest.
In contrast, the elastic $ep$ scattering describes nucleon at rest (with energy $m_N$) absorbing a virtual photon 
($p_{\gamma}=(E_{e,f}-E_{e,i},\vec{p}_{\gamma})$) and starting to move with a small velocity. When $p_{\gamma}^2$ goes to zero, the 
Fourier transformation of the spacelike EMFFs are converted to the electromagnetic (EM) distributions of the static nucleon. 

For the timelike region, we take the vacuum polarization of an electron as an example, where one only needs to change the nucleon into the electron for Fig.A.  
When probed closely, the electron will polarize the vacuum around, known as charge screening.  
Specifically, as shown by the shadow vertex in Fig.A, the ingoing electron emit photons, and the photons can be annihilated (polarized) into electron-anti-electron pairs (and less quark-anti-quark pairs), but with more anti-electrons towards themselves and more electrons forward. 
In the rest-frame of the ingoing electron, one will find that it emits a virtual photon (timelike if the polarized electron-positrons appear) and has negative energy for the outgoing electron. 
The negative energy will be temporary as the probe (for instance, a photon) will transfer energy to the outgoing electron and finally the energy is positive.  
These considerations are quite similar to that of the nucleon in the timelike region as discussed above.  
If one performs the Fourier transformation of Eq.~(10) on the EMFFs of the electron, the momentum of the photon has been 
changed into the distance between the photon and the electron. This $r$ can also be recognized as the distance between the electron and the polarized electric charges,  as the photon will be annihilated into the polarized electron-positron pairs. At the end, one obtains the distributions of the polarized electric charges for the electron.

Going back to the EMFFs of the nucleons in the timelike region, it is realized that the EMFFs indicate that the nucleon (at rest) emits a virtual photon, and the photon will be annihilated into polarized electron-positron pairs, quark-anti-quark pairs, etc.  
The electron-positron pairs would dominate as they are much lighter than other  charged particles. 
By performing Fourier transformation in Eq.~(10), one obtains $r$ as the distance between the polarized particles and the nucleon.
Further, considering that a nucleon is composed of clouds of quarks and gluons, the virtual photon can be emitted from the valence quarks directly, where the EM interactions would dominate. 
Also, it can be emitted from the sea quarks where both EM and strong interactions are involved. Therefore, both EM and strong interactions will affect the distributions of the polarized electric charges.
In addition, the photons emitted by the nucleon are limited to the hard photons ($E_{\gamma^*}>2 m_N$).  
At the end of the day, one concludes that the 
Fourier transformation on the EMFFs of the nucleons in the timelike region, as given in Eq.~(10), represents the distributions of the 
polarized electric charges generated by hard photons for the nucleon. More specifically, the dipole, underdamped oscillation, and overdamped oscillation are for the leading order, next-to-leading order, and next-to-next-to-leading order polarizations. 

\begin{figure}[htp]
    \centering
    \includegraphics[width=0.6\linewidth,height=0.10\textheight]{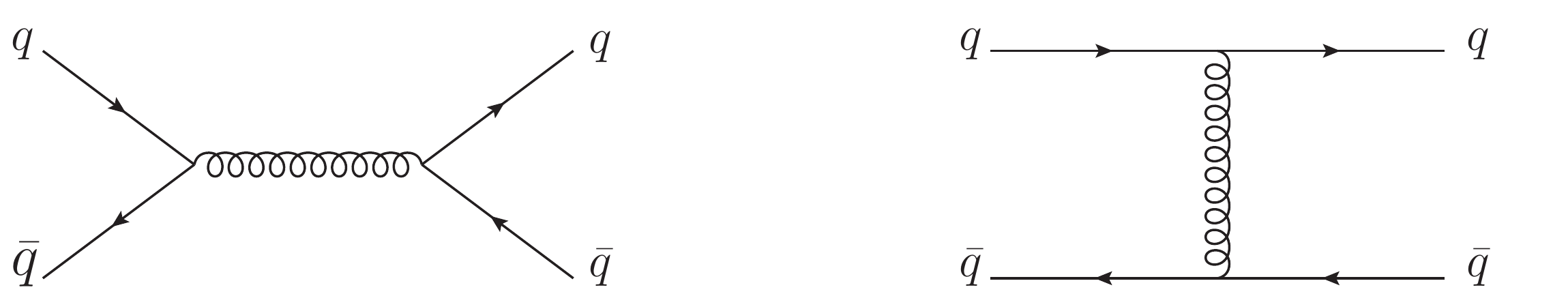}
    \caption{The Feynman diagram of quark interaction in nucleon anti-nucleon scattering}
    \label{fig:quarkFeynman}
\end{figure}

\sectitle{The origin of the oscillations}
As is well known, it is difficult to calculate the $N\bar{N}$ interaction in perturbative QCD. However, some QCD-based phenomenological models can be applied to explore the underlying physics and to shed light on the question of how to generate 
the fractional oscillation. 
The strategy is as follows: we construct a whole potential of $N\bar{N}$ scattering, where the low energy part is taken from 
ChEFT as given in the main text, while the high-energy part is taken from the one-gluon exchange (OGE) within a constituent quark model \cite{Godfrey:1985xj,Lu:2020rog}
\begin{align}
    \label{Eq:NN_H}
    V_{\mathrm{eff}}(\vec{r})=&d\sum_{i=1}^{3}\sum_{j=4}^{6}(G^{ij}_{\mathrm{eff}}(\vec{r})+S^{ij}_{\mathrm{eff}}(\vec{r}))\,,\nonumber\\
    G^{ij}_{\mathrm{eff}}(\vec{r})=&\left(1+\frac{p^2}{E_iE_j}\right)^{1/2}\tilde{G}(r)\left(1+\frac{p^2}{E_iE_j}\right)^{1/2}+\delta_{ij}^{1/2+\epsilon_{\mathrm{so}(v)}}\frac{3(\vec{S}_i+\vec{S}_j)\cdot \vec{L}}{2m^2}\frac{1}{r}\frac{\partial \tilde{G}(r)}{\partial r}\delta_{ij}^{1/2+\epsilon_{\mathrm{so}(v)}}\nonumber\\
    &+\delta_{ij}^{1/2+\epsilon_c}\frac{2\vec{S}_i\cdot\vec{S}_j}{3m^2}\nabla^2\tilde{G}(r)\delta_{ij}^{1/2+\epsilon_c}-\delta_{ij}^{1/2+\epsilon_{t}}\frac{3\vec{S}_i\cdot\hat{r}\vec{S}_j\cdot\hat{r}-\vec{S}_i\cdot\vec{S}_j}{3m^2}\left(\frac{\partial^2}{\partial r^2}-\frac{1}{r}\frac{\partial}{\partial r}\right)\tilde{G}(r)\delta_{ij}^{1/2+\epsilon_{t}}\,,\nonumber\\
    S^{ij}_{\mathrm{eff}}(\vec{r})=&\tilde{S}-\delta_{ij}^{1/2+\epsilon_{\mathrm{so}(s)}}\frac{(\vec{S}_i+\vec{S}_j)\cdot \vec{L}}{2m^2}\frac{1}{r}\frac{\partial \tilde{S}(r)}{\partial r}\delta_{ij}^{1/2+\epsilon_{\mathrm{so}(s)}} \,.
    \tag{B1}
\end{align}
where the subscripts $i$ and $j$ represent quarks and anti-quarks in the nucleons/anti-nucleons, respectively. $E_{i,j},\,\vec{S}_{i,j}$ are energy and spin operators, respectively. $\vec{L}$ is the orbital angular momentum between the quark and the anti-quark. The factor $d$ describes the polarization of the quark. $\tilde{G}$ and $\tilde{S}$ are the smeared results of the Coulomb and linear confinement potential, which are given as 
\begin{align}
\label{eq:GS}
    \tilde{G}=&\vec{F}_i\cdot\vec{F}_j\sum_{k=1}^3\frac{\alpha_k}{r}\mathrm{erf}(\tau_{kij}r)\,,\nonumber\\
    \tilde{S}=&-\frac{3}{4}\vec{F}_i\cdot\vec{F}_j\left\{br\left[\frac{e^{-\sigma_{ij}^2r^2}}{\sqrt{\pi}\sigma_{ij}}+\left(1+\frac{1}{2\sigma_{ij}^2r^2}\mathrm{erf}(\sigma_{ij}r)\right)\right]+c\right\} \,. \tag{B2}
\end{align}
The factors $\delta_{ij}^{1/2+\epsilon_k}$ in Eq.(\ref{Eq:NN_H}) are given as 
\begin{eqnarray}\label{eq:deltaij}
    \delta_{ij}^{1/2+\epsilon_k}=\left(\frac{m^2}{E_iE_j}\right)^{1/2+\epsilon_k} \,, \nonumber
\end{eqnarray}
where $k=c,\,t,\,\mathrm{so}(v),\,\mathrm{so}(s)$, i.e., contact, tensor, vector spin-orbit, and scalar spin-orbit. 
The typical Feynmann diagrams of the OGE are shown in Fig.\ref{fig:quarkFeynman}. 
We have assumed that the OGEs happen only between the quarks in the nucleon and the antiquarks in the antinucleon.
The OGE of two quarks or antiquarks in the nucleon or antinucleon is not considered. 
Also, OGE is for a pair of quark-antiquark only, while the other two quarks/antiquarks are observators and do not interact simultaneously.   
Notice that the expected value of the color matrix elements for the $s-$channel exchanges are given as $\langle \vec{F}_i\cdot\vec{F}_j\rangle=-\frac{1}{9}$, while for the $t$-channel exchanges, one has $\langle \vec{F}_i\cdot\vec{F}_j\rangle=0$. 
Only $s$ channels of OGEs contribute, and there are five Feynman diagrams for $p\bar{p}\to p\bar{p}$ and $n\bar{n}\to n\bar{n}$, and four Feynman diagrams for $p\bar{p}\leftrightarrow n\bar{n}$. 

With the constructed potentials, we need to transform them from the coordinate space to the momentum space to solve the Lippmann-Schwinger equation. The Fourier transformation is
\begin{align}
\label{eq:VrtoVq}
    V_{\mathrm{OGE}}(\vec{q})=\int\frac{d^3r}{(2\pi)^3}e^{-i\vec{q}\cdot\vec{r}}V_{\mathrm{OGE}}(\vec{r})\,, \tag{B3}
\end{align}
where $q=|\vec{p}'-\vec{p}|$. The partial wave projection is also performed in the potential to get the information of the ${}^3S_1-{}^3D_1$ coupled channel scatterings. 
Notice that the OGE potential is of short-distance range, so we use the upper limit of $r_c=$1.5~fm for the radius ($\int_0^{r_c}dr$) to integrate Eq.~\eqref{eq:VrtoVq}. 
The ChEFT potential dominates the low-energy region, with the cut-off $\Lambda=850$~MeV (corresponding to $\sqrt{s}=2.2$~GeV) as used in the main text. Meanwhile, the OGE potential $V^{\mathrm{OGE}}$ dominates the high-energy region, i.e., from 2.2~GeV up to 3.2 GeV. Then the total potential have following form
\begin{align}
    V^{\mathrm{tot}}(p,p')=V^{\mathrm{ChEFT}}(p,p')f(p,p',\Lambda)+V^{\mathrm{OGE}}(p,p')\tilde{f}(p,p',\Lambda)f(p,p',\Lambda')\,. \tag{B4}
\end{align}
The regulator functions multiplied by the potentials are chosen as \cite{Dai:2017ont}
\begin{align}
f(p,p',\Lambda)=\exp\left(-\frac{p^6+p'^6}{\Lambda^6}\right)\,,\;\;\; \tilde{f}(p,p',\Lambda)=\left[1-\exp\left(-\frac{p^2+p'^2}{\Lambda^2}\right)\right]^4\,,\tag{B5}
\end{align}
where the former is to suppress the high energy interaction and the latter to suppress the low energy interaction. Notice that one has $\Lambda=850$~MeV and $\Lambda'=1500$~MeV, corresponding to the upper limits of our analysis, $\sqrt{s}=$2.2 GeV and 3.2 GeV, respectively. These regulator functions are needed to avoid double counting between the potentials of ChEFT and OGE, where in the low-energy region the scattering has already been well described by ChEFT, and in the high-energy region, the OGE potentials describe the interactions well.
The amplitude $T^{\mathrm{tot}}$, which can describe the energy region from threshold to 3.2 GeV, is obtained by solving the LES in Eq.~(3). Since the high energies (above 2.2~GeV) are considered, the Born term in the DWBA approach needs to be re-fixed. We take the following form
\begin{align}
    f^{0,\bar{N}N}_{L, \mathrm{tot}}(k;E_k)=f^{0,\bar{N}N}_{L, \mathrm{ChEFT}}(k;E_k)f(k,\Lambda)+f^{0,\bar{N}N}_{L, \mathrm{OGE}}(k;E_k)\tilde{f}(k,\Lambda)f(k,\Lambda')\,. \tag{B6}
\end{align}
With the regulator functions $f(k,\Lambda)=\exp\left(-\frac{k^6}{\Lambda^6}\right)$ and $\tilde{f}(k,\Lambda)=\left[1-\exp\left(-\frac{k^2}{\Lambda^2}\right)\right]^4$.

\begin{figure}[htp]
    \centering
    \includegraphics[width=0.8\linewidth,height=0.2\textheight]{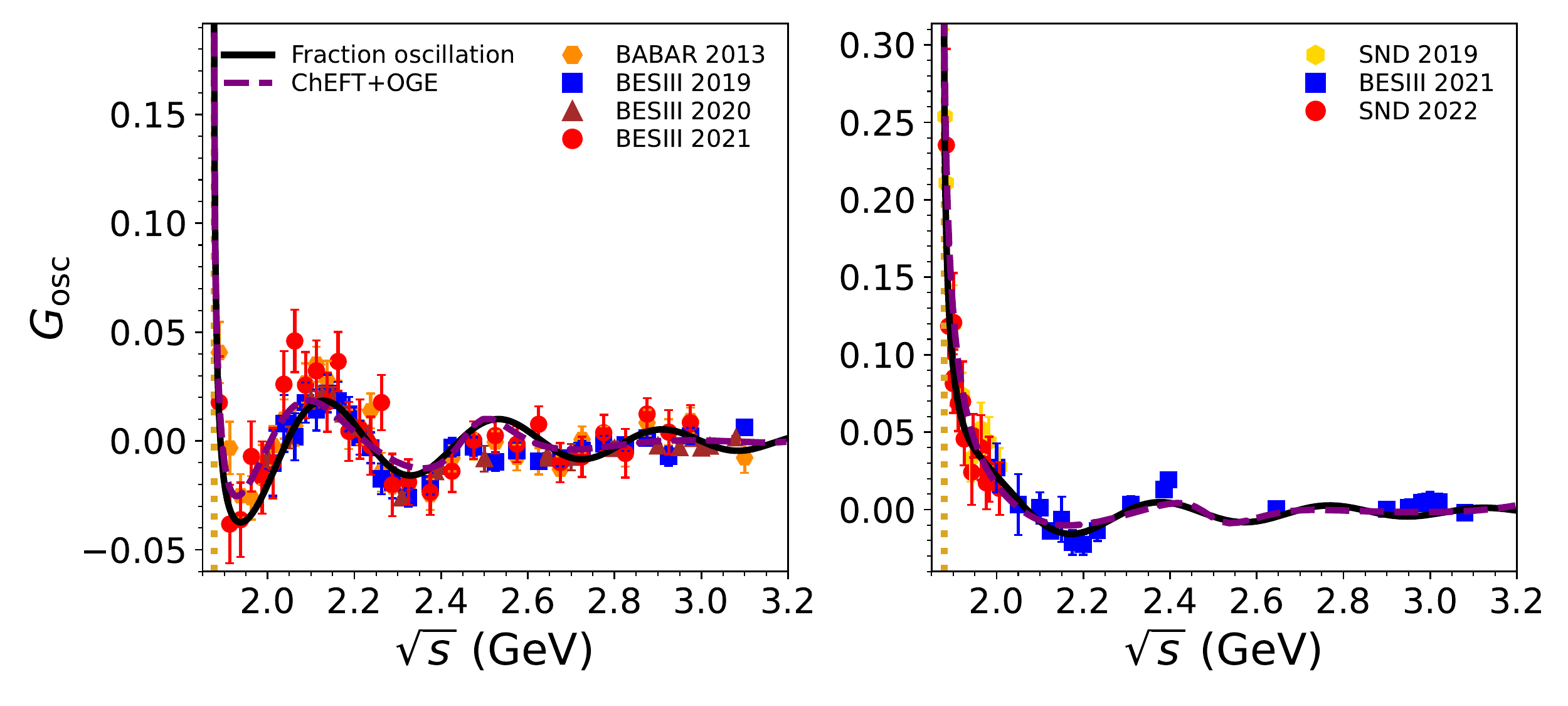}
    \caption{The results for $G_{\mathrm{osc}}$.}
    \label{fig:GoscH}
\end{figure}

These new potentials can describe the data of $e^+e^-\to \bar{N}N$ well, and we extract the $G_{\mathrm{osc}}$, see Fig.\ref{fig:GoscH}. 
With this potential, we can now solve the amplitudes from $N\bar{N}$ threshold up to 3.2~GeV.  And then the DWBA method is used to obtain the $e^+e^-\to N\bar{N}$ amplitude. The cross sections and effective form factor are refitted to fix the parameters, such as $d$ in Eq. \eqref{Eq:NN_H}, $b,\,c$ in Eq.\eqref{eq:GS}, and $\epsilon_{c},\,\epsilon_{t},\,\epsilon_{\mathrm{so}(v)},\,\epsilon_{\mathrm{so}(s)}$ in Eq.\eqref{eq:deltaij}. 
One can see that the combined interactions between the low-energy potential from ChEFT and the high-energy potential from OGE can roughly generate the oscillation behavior of the fractional model.  
Notice that without the one-gluon interaction, there would not be oscillations in the high-energy region. Hence, from it, we recognize that the OGE potential is essential to the oscillation (underdamped) in the high-energy region. In contrast, in the low-energy region, the ChEFT potential dominates the generation of the overdamped oscillation.

\sectitle{SFFs with different background functions}
In the main text, the background is fixed to be a dipole function, i.e., the same as what is used by the experimentalists. However, it would be interesting to consider different forms and check whether the oscillation of the SFFs still exists. The dipole functions for the proton and neutron in the main text are given as \cite{Tomasi-Gustafsson:2001wyw}, 
\begin{align}
    |F^p_{1}(q^2)|=\frac{\mathcal{A}^p}{(1+q^2/m_{a}^{2})(1-q^2/q_0^2)^2},\quad |F^n_{1}(q^2)|=\frac{\mathcal{A}^n}{(1-q^2/q_0^2)^2}\,,\tag{C1}
    \label{eq:dipole;1}
\end{align}
where the parameters are fixed by the experiment \cite{BESIII:2021rqk}, 
$\mathcal{A}^p$=7.7, $\mathcal{A}^n=3.5\pm0.1$, $m_a^2$=14.8 ($\mathrm{GeV}/c$)$^2$ and $q_0^2$=0.71 ($\mathrm{GeV}/c$)$^2$.  

\begin{figure}[htp]
    \centering    \includegraphics[width=0.7\linewidth,height=0.26\textheight]{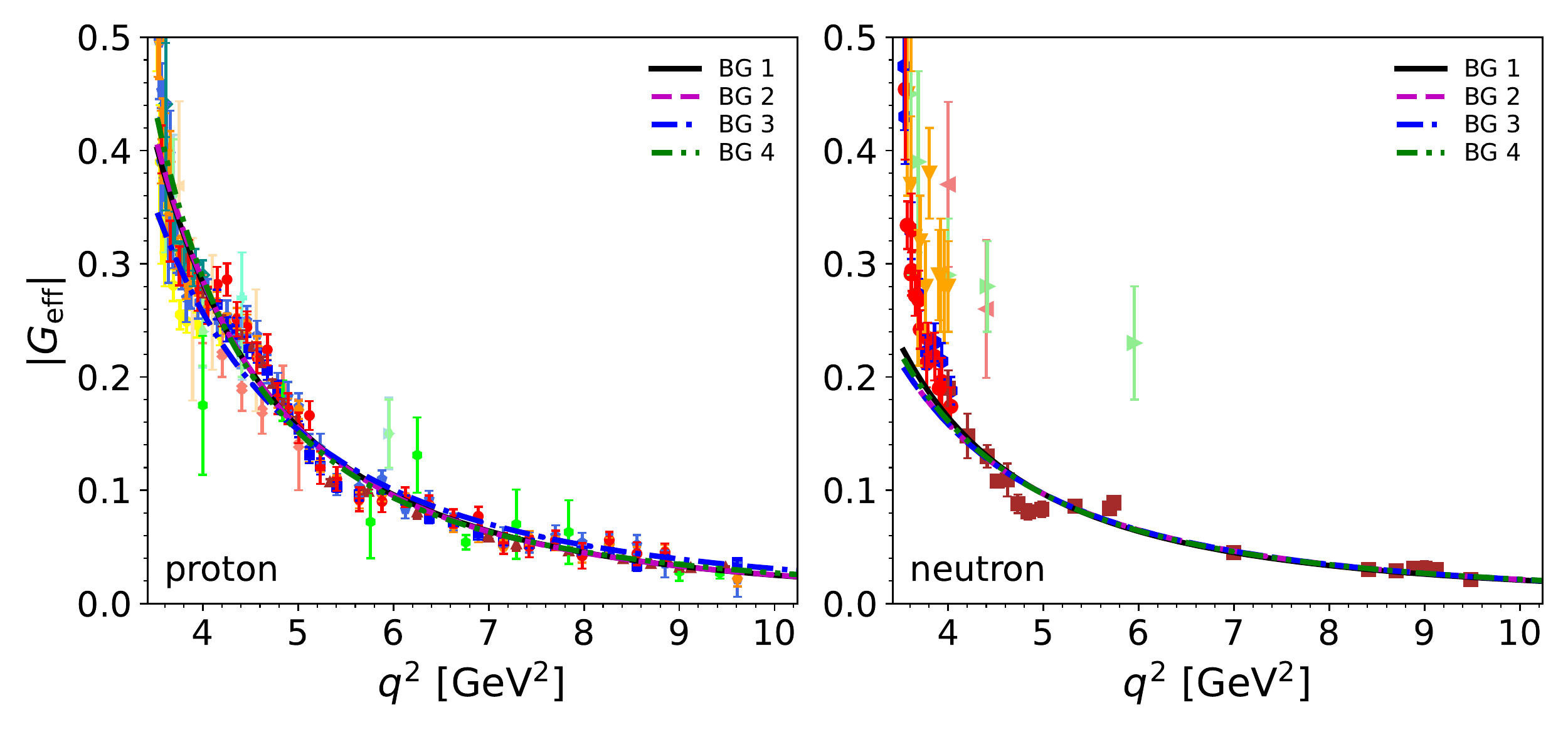}
    \caption{The background contributions to EMFFs of the proton and neutron, with different background functions in Eqs.~(\ref{eq:dipole;1},\ref{eq:dipole;new}). }
    \label{Dipole}
\end{figure}

Since the new datasets are included in this analysis, e.g.,
the data from BESIII \cite{BESIII:2019hdp,BESIII:2019tgo,BESIII:2021rqk} and SND \cite{SND:2022wdb}, we refit the data and obtain the following parameters for the dipole functions,
$\mathcal{A}^p=8.56$, $m_a^2=10.20$~GeV$^2$,  $\mathcal{A}^n=3.53$.  
See the black solid lines in Fig.~\ref{Dipole}, named as \lq BG 1'.

There are other functions in the literature~\cite{Lepage:1979za,E835:1999mlt,Shirkov:1997wi,Brodsky:2007hb,Bianconi:2015vva} that can be considered for 
describing the background. Here we explore some of them and we refit the data to fix the parameter, resulting in different kinds of backgrounds.
The concrete functions considered are 
\begin{align}
    &|F_2(q^2)|=\frac{\mathcal{A}_2^{n,p}}{(q^2)^2\log^2(q^2/\Lambda_{2n,p}^2)}\,, \nonumber\\
    &|F_3(q^2)|=\frac{\mathcal{A}_3^{n,p}}{(q^2)^2[\log^2(q^2/\Lambda_{3n,p}^2)+\pi^2]}\,, \nonumber\\
    &|F_4(q^2)|=\frac{\mathcal{A}_4^{n,p}}{(1-q^2/m_{4p,n}^2)(2-q^2/\tilde{m}_{4p,n}^2)}\,,\tag{C2} \label{eq:dipole;new}
\end{align}
with fitting parameters of $\mathcal{A}_2^{p}=32.1$ GeV$^4$, $\Lambda_{2p}=0.53$ GeV, $\mathcal{A}_2^{n}=235$ GeV$^4$, and $\Lambda_{2n}=0.016$ GeV for \lq BG 2'. See the magenta dashed lines in Fig.~\ref{Dipole}. 
The fitting parameters for \lq BG 3' are given as $\mathcal{A}_3^{p}=72.8$ GeV$^4$, $\Lambda_{3p}=0.49$ GeV, $\mathcal{A}_3^{n}=210$ GeV$^4$, and $\Lambda_{3n}=0.028$ GeV.  
The background contributions are shown by the blue dash-dotted lines in Fig.~\ref{Dipole}. 
The fitting parameters of \lq BG 4' are given as
\begin{eqnarray}
    \mathcal{A}_4^p&=&2.20,\quad m_{4p}^2=1.35\,\mathrm{GeV}^2,\quad \tilde{m}_{4p}^2=0.68\,\mathrm{GeV}^2,\nonumber\\
    \mathcal{A}_4^n&=&11.48,\quad m_{4n}^2=0.57\,\mathrm{GeV}^2,\quad \tilde{m}_{4n}^2=0.29\,\mathrm{GeV}^2. \nonumber
\end{eqnarray}
See the green dash-dot-dotted lines in Fig. \ref{Dipole}. 

\begin{figure}[htp]
    \centering \includegraphics[width=0.98\linewidth,height=0.5\textheight]{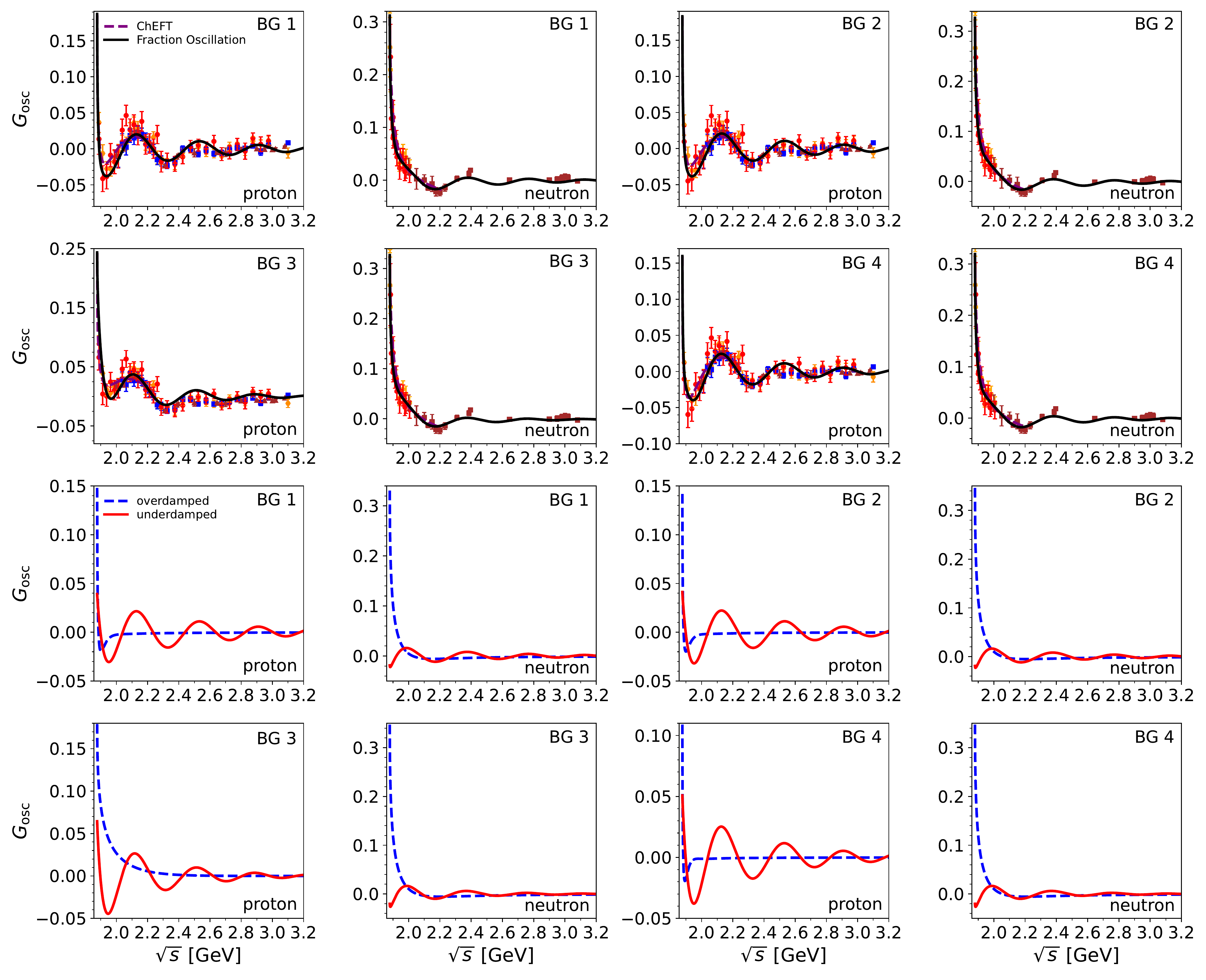}
    \caption{Results for the fractional oscillators fitted to the SFFs, considering different functions for the background.}
    \label{GoscDiffDipole}
\end{figure}

To test the stability of the oscillation, we fit our fractional oscillators to the SFFs subtracted by the four background functions discussed above. 
The results are shown in Fig. \ref{GoscDiffDipole}. 
The purple dashed lines are the SFFs calculated by ChEFT in the low energy region. The black solid lines are the new fitting results for the fractional oscillators.  One can see that the fractional oscillations are still obvious, and they describe the experimental data and the results of ChEFT well. This indicates that our fractional oscillation model is stable.

\end{document}